\documentclass[eqsecnum,aps,epsfig]{revtex4}
\usepackage{amsmath,amssymb,amsthm,bm}
\usepackage{psfrag}
\usepackage[dvips]{graphicx}

\begin{document}
\title{Confinement and the second vortex of the SU(4) gauge group}
\author{S.~Deldar$^\dagger$ and S.~Rafibakhsh$^\ddagger$}
\affiliation{
Department of Physics, University of Tehran, P.O. Box 14395/547, Tehran 1439955961,
Iran \\
$^\dagger$E-mail: sdeldar@ut.ac.ir\\
 $^\ddagger$E-mail: rafibakhsh@ut.ac.ir
}
\date{\today}

\begin{abstract}

We study the potential between static SU(4) sources using the Model of Thick Center
Vortices. Such vortices are characterized by the center elements $z_1=\mathrm i$ and
$z_2=z_1^2$. Fitting the ratios of string tensions to those obtained in Monte-Carlo
calculations of lattice QCD we get $f_2>f_1^2$, where $f_n$ is the probability that
a vortex of type $n$ is piercing a plaquette. Because of $z_2=z_1^2$ vortices of
type two are overlapping vortices of type one. Therefore, $f_2>f_1^2$ corresponds to
the existence of an attractive force between vortices of type one.   
\end{abstract}

\maketitle

\section{INTRODUCTION}\label{Intro}

It is a well known fact that Quantum Chromodynamics (QCD) is truly described by the
SU(3) gauge group. However, the mechanism of confinement is still under intensive
discussion and is an attractive open subject. Studying the SU(N) gauge theories with
$N$ greater than three, helps us to better understand different aspects of QCD.
Confinement implies that the potential between static sources has a linear term 
\begin{equation}
V \simeq \sigma R 
\end{equation}
where $\sigma$ is the string tension and $R$ is the distance between the two
sources. SU(N) gauge theories with $N>3$ have some characteristics that SU(3) does
not have. For example, there exists only one universal string tension for the SU(3)
gauge theory but for SU(N) with $N>3$, the number of independent stable string
tensions are equal to $\mathrm{int}[\frac{N}{2}]$. The values of these string
tensions will constrain the details of the confinement models. Both meta-stable
string tensions which are the string tensions at intermediate distances and the
stable string tensions  which are the string tensions at large distances, are
important, in this respect. Stable string tensions depend on the N-ality $k$ of the
corresponding representation. In fact, at large distances where the distance between
two sources increases, a pair of gluons pops out of the vacuum and couples with the
initial sources. They do not change the N-ality of the representation but reduce the
dimension of 
 the representation to the lowest dimension of the corresponding N-ality. Therefore,
stable (asymptotic) string tensions depend on the N-ality of the representation. On
the other hand, at intermediate distances, the string tensions depend on the
dimension of the representation. For this region, the potential energy is not large
enough to produce a pair of gluons and therefore, the dimension of the original
sources does not change. 

For stable strings, there are some existing theories describing the ratio of
$\frac{\sigma_{k}}{\sigma_\mathrm{f}}$ where $\sigma_{\mathrm{f}}$ and
${\sigma_{k}}$ are the string tensions for fundamental quarks and for quarks of
representations with N-ality $k$, respectively. The linear potential between two 
static sources may be explained by forming a chromoelectric flux tube carrying
charge in the center $Z_{N}$ of the gauge group. The most trivial idea is that the
total flux is carried by $k$ independent fundamental tubes. Then
\begin{equation}
\sigma_{k}=\tilde{k}\sigma_\mathrm{f}\qquad ;\qquad \tilde{k}=\min\{k,N-k\}
\label{th1}
\end{equation}
Because of charge conjugation we get $\sigma_{k}=\sigma_{N-k}$. Thus, for the SU(3)
gauge group, $\sigma_2=\sigma_1$ and one universal string tension is obtained. If
one wants to study theories with more than one string tension, one has to study
SU(N) with $N>3$. The asymptotic Casimir scaling is another theory which claims that
\cite{Ambjorn:1984}
\begin{equation}
\frac{\sigma_{k}}{\sigma_\mathrm{f}}=\frac{k(N-k)}{N-1}
\label{th2}
\end{equation}
Calculations in brane M-theory \cite{Douglas:1995nw}, predict Sine-law scaling
\begin{equation}
\frac{\sigma_{k}}{\sigma_\mathrm{f}}=\frac{\sin\frac{k\pi}{N}}{\sin\frac{\pi}{N}}
\label{th3}
\end{equation}

On the other hand, for intermediate distances, lattice calculations show that the
string tensions are roughly proportional to the eigenvalues of the quadratic Casimir
operators \cite{Deldar:1999vi}. In ref.~\cite{DelDebbio:1995gc} this phenomenon was
dubbed ``Casimir scaling''. The Casimir scaling regime is expected to extend roughly
from the onset of confinement to the onset of screening \cite{Faber:1997rp}. There
is another argument about the linear part of the potential at intermediate distances
which claims that the string tension in this region is proportional to the number of
fundamental flux tubes embedded into the representation \cite{Michael:1998sm}. The
fundamental flux or string is the one that connects a fundamental heavy quark with
an anti-quark. This idea is called ``flux tube counting''. In general, the
fundamental strings do not interact at very large $N$. Thus if one interprets the
meta stable string tensions between the higher representation sources to be proportional to the number of flux tubes with  $\frac{1}{N}$ corrections; then for
large enough distances these meta-stable strings will decay into stable strings
with given $N$-ality, whose tension is likely to be described by the Sine formula
or by Casimir scaling in some approximation which is not yet known so far.

In this article we study the string tensions of SU(4) static sources, within the
model of thick center vortices. In our previous calculations \cite{Deldar:2004hg},
we have shown that using only the first non-trivial type of the vortices of the
SU(4) gauge group, two different asymptotic string tensions may be obtained, one for
$k=1$ and another one for $k=2$. Their ratio approaches asymptotically
$\frac{\sigma_{k=2}}{\sigma_{k=1}}\simeq 2$ in agreement with flux tube counting,
see Eq.~(\ref{th1}). If one wants to get other asymptotic string ratios, e.g. those
corresponding to lattice calculations, one has to use both vortex types of the SU(4)
gauge group. In the following sections, after giving a brief review of the role of
vortices for the confinement of quarks, potentials between static sources are
calculated using both types of SU(4) vortices. Because of the relation between the
ratio of the stable string tension, $\frac{\sigma_{k}}{\sigma_\mathrm{f}}$, where
$\sigma_\mathrm{f}$ is the fundamental string tension, and the probabilities  $f_1$  and $f_2$
of piercing plaquettes by vortices of type one and two, one can determine the ratio
of the probabilities, $\frac{f_2}{f_1}$, by fixing
$\frac{\sigma_{k}}{\sigma_\mathrm{f}}$ from lattice calculations or from the
theories related to Eqs.~(\ref{th1})-(\ref{th3}). Then, the induced potentials may
be determined from the model. We show that the meta-stable string tension ratio
$\frac{\sigma_{r}}{\sigma_\mathrm{f}}$ is almost independent of the ratio of the
asymptotic string tensions, $\frac{\sigma_{k}}{\sigma_\mathrm{f}}$, where
$\sigma_{r}$ is the string tension of representation $r$ at intermediate distances.
Approximate agreement with both Casimir scaling and flux tube counting is observed
for all potentials at intermediate distances. The effect of the second vortex type which
modifies the concavity of the potentials and  the type of the
interaction between vortices are also discussed.

\section{Potentials for two types of thick center vortices}

The vortex model of QCD assumes that the vacuum of quantum chromodynamic is filled
with vortices of finite thickness which carry magnetic fluxes corresponding to the
center of the gauge group. In order that vortices have finite energy per unit
length, their gauge potential at large transverse distances must be a pure gauge.
The non-trivial nature of gauge transformations for producing the gauge potentials
causes that the vortex cores have non-zero energy and makes vortices topologically
stable. The potential energy between static sources induced by the vortices is
\cite{Faber:1997rp}
\begin{equation}
V(R) = -\sum_{x \in A}\ln\left\{ 1 - \sum^{N-1}_{n=1} f_{n}
(1 - {\mathrm {Re}} {\cal G}_{r} [\vec{\alpha}^n_{C}(x)])\right\}.
\label{VR}
\end{equation}
$x$ is the location of the center of the vortex and $C$ indicates the Wilson loop.
$A$ is the minimal area of the Wilson loop and the sum over positions $x$ runs over
all plaquettes in the plane of the loop. $f_{n}$ represents the probability that any
given unit area is pierced by a vortex of type $n$. ${\cal G}_{r}$ which gives the
information about the flux distribution and the contribution that a vortex with its
center in a specific plaquette may have to the Wilson loop is given by
\begin{equation}
{\cal G}_{r}[\vec{\alpha}] = \frac{1}{d_{r}}
{\mathrm {Tr}} \exp[{\mathrm {i}}\vec{\alpha} . \vec{H}^r],
\label{calG}
\end{equation}
where $d_{r}$ is the dimension of the representation $r$ and
$\{H_i^r,i=1,2,...,N-1\}$ are the generators in this representation spanning the
Cartan subalgebra. The vector $\vec\alpha_{c}(x)$ depends on the fraction of the
vortex core which is enclosed by the loop, thus on the color structure and the
position of the vortex and the shape of the loop. For each SU(N) gauge group, there
are $N-1$ types of vortices, corresponding to the non-trivial center elements $z_n$.
Vortices of type $n$ and $N-n$ differ in the direction of the magnetic flux only and
have the same probability $f_{n}$. Thus
\begin{equation}
f_{n}=f_{N-n}\qquad\mathrm{and}\qquad{\cal G}_{r} [\vec{\alpha}^n_{C}(x)]=
{\cal G}_{r}^\star [\vec{\alpha}^{N-n}_{C}(x)].
\label{fequal}
\end{equation}
Hence, among the three vortices of the SU(4) gauge group, two of them may be
considered to be the same. Therefore, $f_1$ is equal to $f_3$ and
${\mathrm {Re}} {\cal G}_{r} [\vec{\alpha}^{(1)}_{C}(x)]=
{\mathrm {Re}} {\cal G}_{r} [\vec{\alpha}^{(3)}_{C}(x)]$, where the upper index
$(n)$ in $\vec{\alpha}^{(n)}_{C}(x)$ indicates th{e type of the vortex.

According to Eq.~(\ref{VR}) the induced potential between SU(4) sources may be
written as
\begin{equation}
V(R)=-\sum_{x} \ln\left\{1-2f_1\left(1 - {\mathrm {Re}} {\cal
G}_{r}\left[\vec{\alpha}^{(1)}_{C}(x)\right]\right)-f_2\left(1 - {\mathrm {Re}}
{\cal
G}_{r}\left[\vec{\alpha}^{(2)}_{C}(x)\right]\right)\right\}.
\label{VR2}
\end{equation}
Since $f_1=f_3$, the first term in equation (\ref{VR2}) is multiplied by 2. To
determine ${\cal G}_{r}\left[\vec{\alpha}\right]$ from equation (\ref{calG}), $H$'s
and $\vec\alpha$'s for the SU(4) gauge group should be determined. The  defining
(fundamental) representation of the three generators $H_i$ of the Cartan sub-algebra
may be chosen as
\begin{eqnarray}
\begin{aligned}
H_1&=\frac{1}{2}(1,-1,0,0),\\
H_2&=\frac{1}{2\sqrt{3}}(1,1,-2,0) ,\\
H_3&=\frac{1}{2\sqrt{6}}(1,1,1,-3).
\label{H}
\end{aligned}
\end{eqnarray}

A center vortex completely linked to a Wilson loop, in the fundamental
representation of the SU(N) gauge group, has the effect of multiplying the Wilson
loop by a non-trivial center element $z_n=\exp\{\frac{2\mathrm i\pi n}{N}\}$
\begin{equation}
W(C)\rightarrow \exp\{\frac{2\mathrm i\pi n}{N}\}W(C) \qquad
n=1, 2, \cdots, N-1.
\end{equation}
For the group SU(4), a vortex of type $n=1$ requires $z_1=\mathrm i$ and a vortex of
type $n=2$ requests $z_2=-1$. The percentage of linking we describe  by a function
$\eta(x)$ which is zero when the vortex does not touch the loop and equal to one if
the vortex is entirely contained within the loop. Then, the influence of a vortex of
type $n$ on the Wilson loop is
\begin{equation}
\mathcal{G}_r[\vec\alpha^{(n)}]=\frac{1}{4}\mathrm{Tr}\exp\{\mathrm
i\vec\alpha^{(n)}(x)\vec H\},
\qquad \textrm{with}\qquad\vec\alpha^{(n)}(x)=\alpha_\mathrm{max}^{(n)}\vec e\,\eta(x)
\end{equation}
A possible choice for the unit vector $\vec e$ and $\alpha_\mathrm{max}^{(n)}$ is
\begin{equation}
\vec e= (0,0,1),\quad\alpha_\mathrm{max}^{(1)}=\pi\sqrt{6},
\quad\alpha_\mathrm{max}^{(2)}=2\pi\sqrt{6},
\label{maxvalues}
\end{equation}
leading to
\begin{equation}
\mathcal{G}_r[\vec\alpha^{(n)}]
=\frac{1}{4}\mathrm{Tr}\exp\{\mathrm i\eta(x)n\pi\sqrt{6}H_3\}
=\frac{1}{4}[3\exp\{\mathrm i\eta(x)n\frac{\pi}{2}\}
 +\exp\{\mathrm i\eta(x)n\frac{3\pi}{2}\}]
\label{GRA}
\end{equation}
We can choose other unit vectors which lead to a permutation of the matrix elements
of $H_3$ or to a sign change of $H_3$ and leave the sum in Eq.~(\ref{GRA})
unchanged, e.g.
\begin{equation}
\vec e= (\sqrt{\frac{2}{3}},\frac{\sqrt{2}}{3},\frac{1}{3})\qquad
\mathrm{with}\qquad
\vec e\vec H=\frac{1}{2\sqrt{6}}\mathrm{diag}(3,-1,-1,-1)
\end{equation}

We have used the following profile, see ref.~\cite{Faber:1997rp},
\begin{equation}
\eta(x)=\frac{1}{2}[1-\tanh(ay(x)+\frac{b}{R})],
\label{profile}
\end{equation}
where
\begin{equation}
y(x)=\begin{cases}
x-R &\quad\mathrm{for}\quad|R-x|\le|x|\\
-x  &\quad\mathrm{for}\quad|R-x|>  |x|\end{cases}
\end{equation}
is the distance of the vortex center to the nearest timelike side of the loop. The
parameter $a$ is of the order of the inverse of the vortex core
thickness~\cite{Faber:1997rp} and the parameter $b$ takes into account that vortices
can not be completely linked to small Wilson loops. 

Therefore, the vectors $\vec{\alpha}^{(n)}(x)$satisfy the following conditions
\begin{enumerate}
\item Vortices which pierce the plane far outside the loop do not affect the loop.
It means for fixed $R$, as $x\to \infty$, $\alpha\to 0$.
\item If the vortex core is entirely contained within the Wilson loop, then for
vortex type one, $|\vec{\alpha}^{(1)}|=\sqrt{6}\pi$ and for vortex type two,
$|\vec{\alpha}^{(2)}|=2\sqrt{6}\pi$.
\item As $R\to 0$ then $|\alpha^{(n)}|\to 0$.
\end{enumerate}
So far, we have determined the group valued functions ${\cal
G}_{r}\left[\vec{\alpha}\right]$ of equation (\ref{calG}). However, the two
parameters $f_1$ and $f_2$ are not specified yet. In the next section, we show that
the asymptotic string tensions depend on the 4-ality class $k$ of the Wilson loop
and deduce from the asymptotic string tension ratios the probability ratios
$\frac{f_2}{f_1}$.

\section{Asymptotic string tensions}

The asymptotic string tensions can be determined from very large Wilson loops. In
this regime, we can neglect the finite thickness of vortices and assume that
vortices piercing the minimal area of the loop would insert a center element
somewhere in the product of link variables. For the SU(4) gauge group, there are
four center elements
\begin{equation}
z_0=1\qquad z_1=\exp{(\frac{\pi\mathrm i}{2})}\qquad 
z_2=\exp{(\pi\mathrm i)}\qquad z_3=z^\ast_1=\exp{(\frac{3\pi \mathrm i}{2})}.
\label{z-center}
\end{equation}
Which center element contributes depends on the type $n$ of the vortex and the
N-ality $k$ of the representation $r$ of the loop. The Young-tableaux of the lowest
representations of 4-ality $k$ for SU(4) and their dimensions are depicted in
Fig.~\ref{young}.
\begin{figure}[h]
\psfrag{k=1}{$k=1$}\psfrag{k=2}{$k=2$}\psfrag{k=-1}{$k=-1$}\psfrag{k=0}{$k=0$}
\psfrag{4}{$4$}\psfrag{6}{$6$}\psfrag{10}{$10$}
\psfrag{20s}{$20_\mathrm{s}$}\psfrag{20ms}{$20_\mathrm{ms}$}\psfrag{4b}{$\bar 4$}
\psfrag{35}{$35$}\psfrag{45}{$45$}\psfrag{20b}{$20_\Box$}
\psfrag{15}{$15$}\psfrag{1}{$1$}
\centering
\includegraphics[scale=0.4]{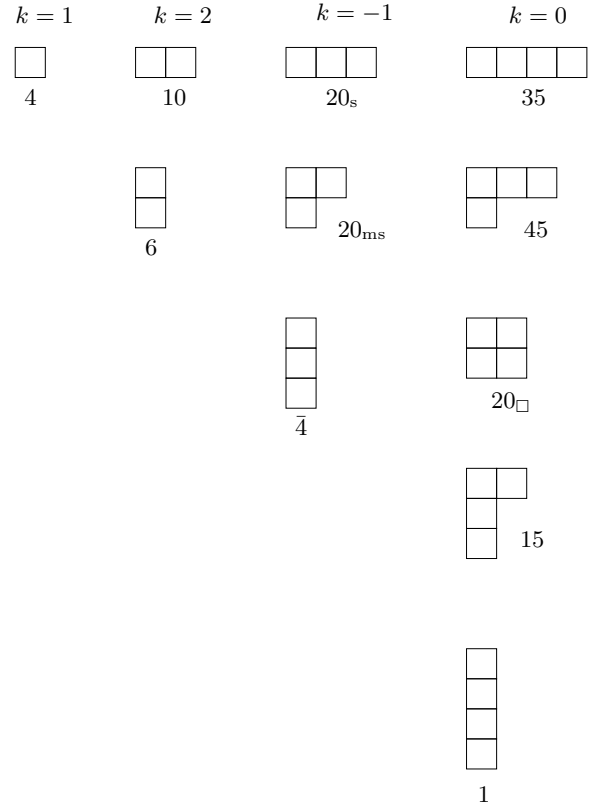}
\caption{The Young tableaus for SU(4) representations of various dimensions and
their N-ality $k$ is shown up to four-quark states.}
\label{young}
\end{figure}
One can easily verify in this figure that the N-ality $k$ modulo $N$ is given by the
number of quarks which is equal to the number of squares in the Young tableau which
is necessary to get such a representation. A representation $r$ contributes with a
factor $\mathcal{G}_r=z_n^{k(r)}$. Hence, from equation (\ref{VR}), the asymptotic
potential for a representation $r$ of N-ality $k$ reads
\begin{equation}
V_r(R)\simeq-\sum_{x\in A}\ln\left\{1 -\sum^{N-1}_{n=1}f_{n}[1 -
Re(z_n^{k(r)})]\right\}
=\exp\{-A\sigma_r\}.
\label{VRR}
\end{equation}

Using $f_1=f_3$ from Eq.~(\ref{fequal}) we get for the fundamental representation of
SU(4)
\begin{eqnarray}
\begin{aligned}
\sigma_\mathrm{f}&=-\ln\left\{1-2f_1\left[1-\mathrm{Re}(z_1)\right]
 -f_2\left[1-\mathrm{Re}(z_2)\right]\right\}=\\
&=-\ln\left[1-2f_1-2f_2\right]\simeq 2f_1+2f_2\qquad
\textrm{for}\qquad(f_1,f_2\ll1)
\end{aligned}
\end{eqnarray}
For the diquark sources with 4-ality $k=2$, like the sextet and the decuplet follows
\begin{eqnarray}
\begin{aligned}
\sigma_6=\sigma_{10}
&=-\ln\left\{1-2f_1\left[1-\mathrm{Re}(z_1^2)\right]
 -f_2\left[1-\mathrm{Re}(z_2^2)\right]\right\}=\\
&=-\ln\left(1-4f_1\right)\simeq 4f_1\qquad
\textrm{for}\qquad(f_1,f_2\ll1)
\end{aligned}
\label{screened6and10}
\end{eqnarray}
Finally, we see that the asymptotic string tension of the 4-ality $k=0$
representations, which can be build by sources with the same numbers of quarks and
antiquarks, is equal to zero. The most important member of this class is the adjoint
representation
\begin{eqnarray}
\sigma_\mathrm{adj}=-\ln\left\{1-2f_1\left[1-\mathrm{Re}(z_1^0)\right]
 -f_2\left[1-\mathrm{Re}(z_2^0)\right]\right\}=0.
\end{eqnarray}
To summarize, the SU(4) asymptotic string tensions behave for small piercing
probabilities like
\begin{eqnarray}
\sigma_\mathrm{f}=2f_1+2f_2,\qquad
\sigma_6=4f_1,\qquad
\sigma_\mathrm{adj}=0.
\end{eqnarray}
This leads to the ratio
\begin{equation}
\frac{\sigma_6}{\sigma_\mathrm{f}}=\frac{2f_1}{f_1+f_2}
\label{frac6f}
\end{equation}
of the asymptotic string tension for representations with 4-ality $k=2$ and 4-ality
$k=1$.

This behavior that the asymptotic string tensions are equal for all quark sources of
the same N-ality one can picture as pairs of gluons which pop out of the vacuum and
screen the original sources. Therefore, the SU(4) gauge group has two universal
asymptotic string tensions and Eq.~(\ref{frac6f}) gives the ratio of these string
tensions within the model of thick center vortices in terms of $f_1$ and $f_2$, of
the probabilities of piercing plaquettes by vortices of type $n=1$ and $n=2$. 

In the next section, we use the ratio of $\frac{\sigma_6}{\sigma_\mathrm{f}}$ from
lattice calculations and from the predictions~(\ref{th1})-(\ref{th3}) to fix the
ratio of $\frac{f_2}{f_1}$ by equation (\ref{frac6f}). Then the potentials between
static sources are calculated and discussed.

\section{Results and Discussion}

In our previous calculations \cite{Deldar:2004hg} for the gauge group SU(4), we
assumed that vortices of type $n=1$ pierce plaquettes with the probability $f_1$ and
no vortices of type $n=2$ are present, $f_2=0$. From Eq.~(\ref{frac6f}) we get under
these assumptions $\sigma_6\simeq 2\sigma_\mathrm{f}$. This agrees with
Eq.~(\ref{th1}) and corresponds to the picture that a string of N-ality $k=2$ is
built from two non-interacting fundamental strings. Lattice calculations do not
confirm this scenario, since they predict
\begin{eqnarray}
\begin{aligned}
\textrm{B.~Lucini, \emph{et. al} \cite{Lucini:2004qp}:}\quad
 &\frac{\sigma_6}{\sigma_\mathrm{f}} \simeq 1.370(20)&\quad\longrightarrow\quad
 \frac{f_2}{f_1}=0.0460(7),\\
\textrm{L. Del Debbio \emph{et. al.} \cite{Deld:2002kp}:}\quad
 &\frac{\sigma_6}{\sigma_\mathrm{f}}\simeq 1.403(15)&\quad\longrightarrow\quad
 \frac{f_2}{f_1}=0.0426(5).
\end{aligned}
\label{LucOht}
\end{eqnarray}
On the other hand, the limit $f_1=f_2$ leads to $\sigma_6=\sigma_\mathrm{f}$, in
obvious contradiction to the lattice results.

\begin{table}
\begin{center}
\begin{tabular}{|l|c|c|c|c|c|c|}\hline
\qquad ratios & $\frac{\sigma_6}{\sigma_\mathrm{f}}$ &
$\frac{\sigma_{15}}{\sigma_\mathrm{f}}$ &$\frac{\sigma_{10}}{\sigma_\mathrm{f}}$
&$\frac{\sigma_{20_{s}}}{\sigma_\mathrm{f}}$
&$\frac{\sigma_{35}}{\sigma_\mathrm{f}}$ &
$\frac{\sigma_6}{\sigma_\mathrm{f}}$(asymptotic) \\
\hline
$f_1=0.1$, $f_2=0$ & 1.50(5) & 1.73(6) & 1.88(5) & 2.65(10) & 3.21(27) & 2 \\
\hline
$f_1=0.1$, $f_2=0.0426[5]$ & 1.51(11)[2] & 1.71(9)[2] & 1.88(10)[2] & 2.57(15)[3] &
3.22(18)[3] & 1.403[15] \\
\hline
$f_1=0.1$, $f_2=0.0460[7]$ & 1.50(8)[2] & 1.69(14)[3] & 1.86(8)[3] & 2.52(10)[4] &
3.16(14)[5] & 1.370[20] \\
\hline
$f_1=0.1$, $f_2=0.1$ & 1.53(15) & 1.71(16) & 2.00(15) & 2.72(21) & 3.39(26) & 1 \\
\hline
$f_1=0$,\quad $f_2=0.1$ & 1.46(19)  & 1.48(9) & 1.67(12) & 2.18(17) & 2.64(18) &  0 \\
\hline
$f_1=0.1$, $f_2=0.050$ & 1.43(20) & 1.71(12) & 1.88(13) & 2.61(18) & 3.22(23) & 1.333\\
\hline
 $f_1=0.1$, $f_2=0.042$& 1.37(20) & 1.78(18) & 1.99(19) & 2.71(26) & 3.34(33) & 1.414\\
\hline
Casimir ratio & 1.33 & 2.13 & 2.4 & 4.2 & 6.4 & \\
\hline
no. of fund. fluxes & 2 & 2 & 2 & 3 & 4 & \\
\hline
\end{tabular}
\end{center}
\caption{The string tension ratios $\frac{\sigma_r}{\sigma_\mathrm{f}}$ for some
SU(4) representations $r$ at intermediate distances are shown in the first five
columns and the values of $\frac{\sigma_6}{\sigma_\mathrm{f}}$ in the asymptotic
region in the last column. The first four rows treat the potentials for the piercing
probabilities $f_1=0.1$ and increasing values of $f_2$, for our previous work
\cite{Deldar:2004hg} in the first row, for the fits to the lattice calculations of
refs.~\cite{Deld:2002kp} and\cite{Lucini:2004qp}  in the second and third row and
for the case $f_1=f_2$ in the forth row. Line five shows the ratios in the absence
of vortices of type $n=1$, line six fits to the predictions of the Casimir scaling
law in Eq.~(\protect\ref{th2}) and line seven fits to the Sine law scaling in
Eq.~(\protect\ref{th3}) in the asymptotic region. These values can be compared with
the the ratio of eigenvalues $ \frac{C_r}{C_\mathrm{f}}$ of the quadratic Casimir
operator i
 n line eight and with the number of fundamental fluxes in line nine. The errors in
square brackets are due to the uncertainties of the lattice data we have used, the
errors in parentheses $\frac{\sigma_r}{\sigma_\mathrm{f}}$ show the systematic
errors due to slight changes of the linear regime.}
\label{TabRatios}
\end{table}
We fix now the ratio of $\frac{f_2}{f_1}$ to the values (\ref{LucOht})
predicted from the lattice data and adjust the absolute values of the
probabilities $f_1$, $f_2$ $a_1$, $a_2$, $b_1$ and $b_2$ such that the
potentials at intermediate distances become linear and get the best 
agreement with the intermediate string tensions of the lattice data. 
For both vortex types, the general form of the vortex profile introduced 
in equation (\ref{profile}) has been used. Fig.~\ref{fig:all} shows 
the potentials of various representations versus $R$ in the range 
of $R\in[1,100]$, using
\begin{eqnarray}
f_1=0.1,\quad f_2=0.046,\quad a_1=0.05,\quad b_1=4.0, \quad a_2=0.025,\quad b_2=8.0.
\end{eqnarray}
Indices one and two refer to vortices of type one and two.
As one can see in Fig.~\ref{fig:all} at large distances, zero 4-ality
representations, like $15$ (adjoint), $20_\Box$ and $35$ are screened. For non-zero
4-ality representations we get two different asymptotic string tensions. The
potentials between the sources with dimension $20_{s}$ and $20_\mathrm{ms}$ become
parallel to that of the fundamental representation and the potentials of the diquark
representations ($6$ and $10$) get the same slope.
\begin{figure}[t]
\centering
\psfrag{f}{f}\psfrag{6}{$6$}\psfrag{10}{$10$}\psfrag{15}{$15$}\psfrag{20s}{$20_\mathrm{s}$}\psfrag{20ms}{$20_\mathrm{ms}$}\psfrag{20b}{$20_\Box$}\psfrag{35}{$35$}
\psfrag{R}{$R$}\psfrag{V(R)}{$V(R)$}
\psfrag{ 0}{$0$}\psfrag{ 10}{$10$}\psfrag{ 20}{$20$}\psfrag{ 30}{$30$}
\psfrag{ 40}{$40$}\psfrag{ 50}{$50$}\psfrag{ 60}{$60$}\psfrag{ 80}{$80$}
\psfrag{ 100}{$100$}
\includegraphics[scale=0.8]{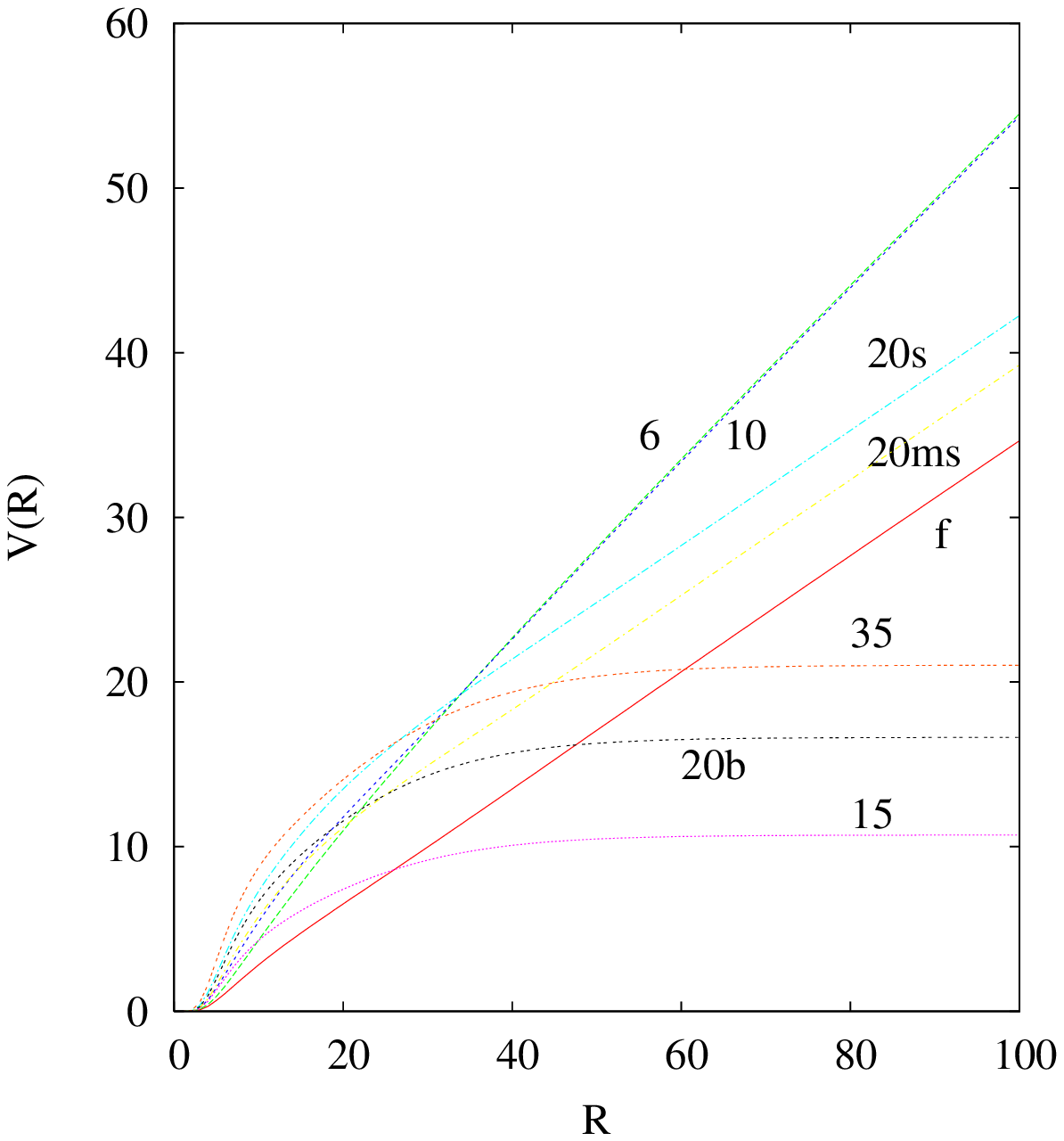}
\caption{Potentials for various representations of SU(4) using both types of
vortices for piercing probabilities $f_1=0.1$ and $f_2=0.046$, see
Eq.~(\protect\ref{LucOht}). The dimensions of the representations are indicated.}
\label{fig:all}
\end{figure}

As expected, we find for all the potentials in Fig.~\ref{fig:all} a linear region at
intermediate distances. Table~\ref{TabRatios} shows the ratios
$\frac{\sigma_r}{\sigma_\mathrm{f}}$ for the corresponding intermediate string
tensions $\sigma_r$ of some of these representations for $f_1=0.1$ and for some
values of $f_2$. The errors in parentheses $\frac{\sigma_r}{\sigma_\mathrm{f}}$ show
the systematic errors due to slight changes of the linear regime. The errors in
square brackets are due to the uncertainties of the lattice data we have used. The
first row contains the results of our previous work \cite{Deldar:2004hg} with
$f_2=0$, the second and the third line the above discussed results from the
comparison with the lattice calculations of refs.~\cite{Deld:2002kp} and 
\cite{Lucini:2004qp} and the forth row the case $f_1=f_2$. It is interesting to
compare these values with the absence of vortices of type $n=1$ in the fifth row
($f_1=0$ and $f_2=0.1$) and with the fits for 
 the asymptotic string tension to the predictions of the Casimir scaling law in
Eq.~(\ref{th2}) and the Sine law scaling in Eq.~(\ref{th3}) in lines six and seven.
It is clearly seen that slight changes of the piercing probabilities $f_2$ have
only a weak influence on the intermediate string tensions. This situation at
intermediate distances differs drastically from the behavior at asymptotic
distances, which was discussed above and is shown for comparison in the last column
of Tab.~\ref{TabRatios}. The asymptotic string tensions of the lattice calculations
in refs.~\cite{Deld:2002kp} and \cite{Lucini:2004qp} can be reproduced only with an
appropriate density of vortices of type $n=2$. The ratio of eigenvalues
$\frac{C_r}{C_\mathrm{f}}$ of the quadratic Casimir operator and finally the number
of fundamental fluxes are represented in lines eight and nine. The lattice results
in Table $17$ of ref.~ \cite{Lucini:2004qp} deviate for the representations $6$ and
$10$ (there is no report on higher dimensional representations) by about $3.5\%$ and $15\%$ from
Casimir scaling, our results in lines 2 and 3 of Table~\ref{TabRatios} by about
$14\%$ and $23\%$. The agreement with Casimir scaling gets worse as the dimensions
of the representations increase. This fact is also observed in
ref.~\cite{Lucini:2004qp} for the representations $6$ and $10$. Considering that
the thick-center-vortices model does not reproduce the Coulombic part, the behavior
of the linear part we get from the model is satisfying.

\begin{figure}[h]
\centering
\psfrag{f1f2}{$f_1=0.1, f_2=0.046$}
\psfrag{f10}{$f_1=0.0, f_2=0.046$}
\psfrag{f20}{$f_1=0.1, f_2=0.0$}
\psfrag{R}{$R$}\psfrag{V(R)}{$V(R)$}
\psfrag{ 0}{$0$}\psfrag{ 10}{$10$}\psfrag{ 20}{$20$}\psfrag{ 30}{$30$}
\psfrag{ 40}{$40$}\psfrag{ 50}{$50$}\psfrag{ 60}{$60$}\psfrag{ 80}{$80$}
\psfrag{ 100}{$100$}
\includegraphics[scale=0.8]{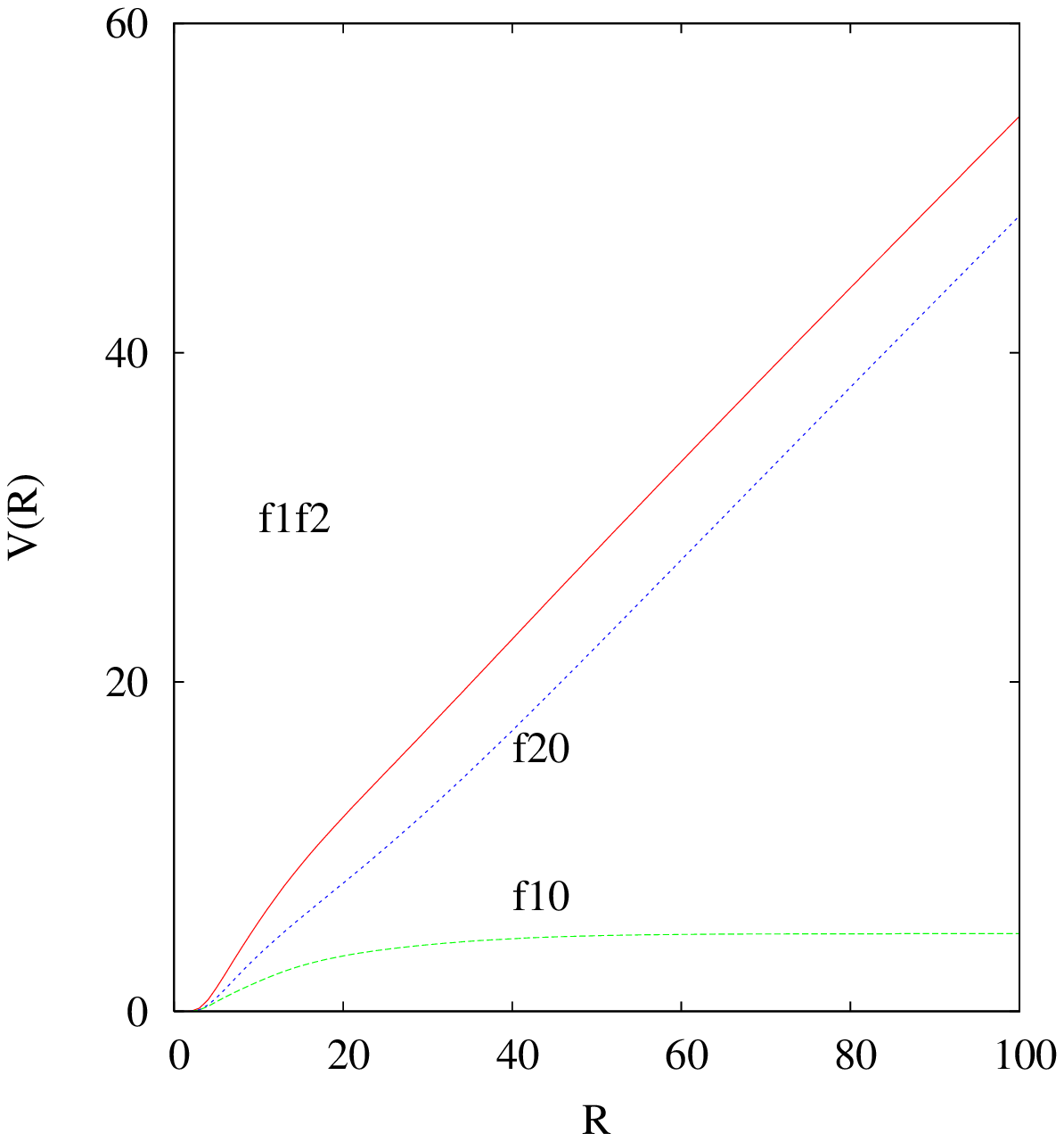}
\caption{For piercing probabilities $f_1$ and $f_2$ small compared to one the
contributions of the two vortex types of SU(4) are almost completely additive as one
can see in these decuplet potentials for the indicated piercing probabilities.
Remarkable is the screening of potentials of 4-ality $k=2$ produced by vortices of
type $n=2$ ($f_1=0.0$ and $f_2=0.046$) which is understandable by a
QCD-Aharanov-Bohm effect.}
\label{fig:gv2_1}
\end{figure}

For values of $f_1$ and $f_2$ which are small enough compared to one, we can expand
the logarithm in Eq.~(\ref{VR2}) around one and separate the contributions for the
potentials produced by the two types of vortices. For our choice of the piercing
probabilities in Fig.~\ref{fig:all}, $f_1=0.1$ and $f_2=0.046$, the violation of
this additivity of the two contributions to the potential by the exact
expression~(\ref{VR2}) is invisible, see Fig.~\ref{fig:gv2_1}, which compares the
full decuplet potential with the potential produced by vortices of type $n=1$
($f_1=0.1$ and $f_2=0.0$) and vortices of type $n=2$ ($f_1=0.0$ and $f_2=0.046$). It
is nice to see that the vortices of type $n=2$ produce $k=2$ potentials which are
asymptotically screened as predicted by Eq.~(\ref{screened6and10}). This can be
understood by the QCD analogon of the Aharanov-Bohm effect. Carrying a two-quark
source around a vortex corresponding to a $z_2=z_1^2=-1$ color magnetic flux leads
to a phase of $z_2^
 2=1$, to a screened potential. It is easily understandable from Eq.~(\ref{VR2})
that the value of this screened potential is roughly proportional to the piercing
probability $f_2$. Due to this screening effect, vortices of type $n=2$ increase
the string tension at intermediate distances only.

There is a probability $f_1^2$ that non-interacting vortices of type $n=1$ have the
same position and due to Eq.~(\ref{maxvalues}) are identified as vortices of type
$n=2$, $f_2=f_1^2$. A comparison to the lattice calculations, see
Eq.~(\ref{LucOht}), leads to $f_2>f_1^2$. This indicates an attraction between
parallel vortex fluxes.

\section{Conclusion}

Confinement is one of the most interesting features of QCD which has been studied by
both lattice gauge theory and phenomenological models. The model of thick center
vortices is one of the phenomenological models which has been fairly successful to
explain the linear part of the potentials. For the SU(N) gauge groups with $N\ge 4$
there exist vortices with different quantized fluxes. In this article we have
studied the gauge group SU(4) which has two types of vortices, vortices of type
$n=1$ with magnetic flux corresponding to the first non-trivial center element
$z_1=\mathrm i$ of SU(4) and vortices of type $n=2$ with a flux corresponding to
$z_2=-1$. We have shown that the ratio $f_2/f_1$ of the probabilities for the
piercing of plaquettes by vortices of both types determines the ratios of asymptotic
string tensions of 4-alities $k=0,1$ and $2$. We underline that the lattice results
for the ratios of these string tension can be explained only by using both types of
vortices
 . Adjusting these ratios to the slightly different results of
refs.~\cite{Deld:2002kp} and \cite{Lucini:2004qp}, the general features of the
potentials and the string tensions at intermediate distances are yet
indistinguishable.

Using vortices of type $n=2$ only, the six and ten dimensional representations,
which have 4-ality $k=2$ and are possible two-quarks states, are screened at large
distances. This results from the multiplication of the Wilson loop holonomy in a
two-quark state by $z_2^2=1$. More generally, it follows that the $k=2$ asymptotic
string tension is independent of the probability $f_2$. The value of $f_2$
influences the $k=\pm 1$ asymptotic string tensions only.

Because of $z_2=z_1^2$ a vortex of type $n=2$ corresponds to two overlapping
vortices of type $n=1$. The analysis of the probabilities $f_1$ and $f_2$ gives an
information about the interaction of vortices. Non-interacting vortices of type
$n=1$ are described by $f_2=f_1^2$, whereas $f_2>f_1^2$ indicates vortex attraction
and $f_2<f_1^2$ vortex repulsion. A comparison of our results with Monte-Carlo
calculations indicates that vortices attract each other.

A consideration of vortices of type $n=2$ modifies the concavity of the potentials
which has been observed in previous calculations of SU(2), SU(3) and SU(4)
potentials with only one type of vortices, see refs.~\cite{Faber:1997rp},
\cite{Deldar:2004hg}, \cite{Deldar:1999yy} and which is not physical
\cite{Bachas:1985xs}. We conjecture that closed unquantised random magnetic flux
lines of rather small size allow to remove this concavity and to introduce a
Coulombic contribution with the correct sign.

\section{Acknowledgments}

We are deeply grateful to numerous discussions with M. Faber. We would also like to
thank 
{\v S}.~Olejn\'{\i}k, J. Greensite and M. Shifman for their valuable comments. This work is partly supported by the research council
 of the University of Tehran and partly by the grant "Structure of matter" 
awarded by the center of excellence to the Department of Physics of the 
University of Tehran.

\end{document}